\documentclass[sigconf]{acmart}

\usepackage{multirow}
\usepackage{multicol}
\usepackage{dblfloatfix}

\makeatletter
\def\blfootnote{\xdef\@thefnmark{}\@footnotetext}
\makeatother

\AtBeginDocument{%
  }

\setcopyright{acmlicensed}
\copyrightyear{2018}
\acmYear{2018}
\acmDOI{XXXXXXX.XXXXXXX}

\acmConference[WiNTECH '24]{The 18th ACM Workshop on Wireless Network
Testbeds, Experimental evaluation \& Characterization 2024}{November 18,
  2024}{Washington, DC, USA}
\acmISBN{978-1-4503-XXXX-X/18/06}

\begin{document}


\title{Cloud-Based Federation Framework and Prototype for Open, Scalable, and Shared Access to NextG and IoT Testbeds}




\author{Maxwell McManus}
\authornote{Both authors contributed equally to this research.}
\email{memcmanu@buffalo.edu}
\orcid{1234-5678-9012}
\affiliation{%
  \institution{University at Buffalo, USA}
  \country{}
}

\author{Tenzin Rinchen}
\authornotemark[1]
\email{tenzinr2@buffalo.edu}
\affiliation{%
  \institution{University at Buffalo, USA}
  \country{}
}

\author{Annoy Dey}
\email{annoydey@buffalo.edu}
\affiliation{%
  \institution{University at Buffalo, USA}
  \country{}
}

\author{Sumanth Thota}
\email{sthota5@buffalo.edu}
\affiliation{%
  \institution{University at Buffalo, USA}
  \country{}
}

\author{Josh (Zhaoxi) Zhang}
\email{zhaoxizh@buffalo.edu}
\affiliation{%
  \institution{University at Buffalo, USA}
  \country{}
}

\author{Jiangqi Hu}
\email{jiangqih@buffalo.edu}
\affiliation{%
  \institution{University at Buffalo, SUNY}
  \country{}
}

\author{Xi (Leo) Wang}
\email{xi.wang1@ufl.edu}
\affiliation{%
  \institution{University of Florida, USA}
  \country{}
}

\author{Mingyue Ji}
\email{mingyueji@ufl.edu}
\affiliation{%
  \institution{University of Florida, USA}
  \country{}
}

\author{Nicholas Mastronarde}
\email{nmastron@buffalo.edu}
\affiliation{%
  \institution{University at Buffalo, USA}
  \country{}
}

\author{Elizabeth Serena Bentley}
\email{elizabeth.bentley.3@us.af.mil}
\affiliation{%
  \institution{U.S. Air Force Research Laboratory}
  \country{}
}

\author{Michael Medley}
\email{michael.medley@sunypoly.edu}
\affiliation{%
  \institution{SUNY Polytechnic Institute, USA}
  \country{}
}

\author{Zhangyu Guan}
\email{guan@buffalo.edu}
\affiliation{%
  \institution{University at Buffalo, USA}
  \country{}
}

\renewcommand{\shortauthors}{McManus et al.}

\begin{abstract}

In this work, we present a new federation framework for \textit{UnionLabs}, an innovative cloud-based resource-sharing infrastructure designed for next-generation (NextG) and Internet of Things (IoT) over-the-air (OTA) experiments. The framework aims to reduce the federation complexity for testbeds developers by automating tedious backend operations, thereby providing scalable federation and remote access to various wireless testbeds. 
We first describe the key components of the new federation framework, including 
the Systems Manager Integration Engine (SMIE),
the Automated Script Generator (ASG), and
the Database Context Manager (DCM).
We then prototype and deploy the new Federation Plane on the Amazon Web Services (AWS) public cloud, demonstrating its effectiveness by federating two wireless testbeds: i) UB NeXT, a 5G-and-beyond (5G+) testbed at the University at Buffalo, and ii) UT IoT, an IoT testbed at the University of Utah\footnote{Mingyue Ji and Xi (Leo) Wang moved from the University of Utah to University of Florida in August 2024.}.  
\blfootnote{This work was supported in part by the National Science Foundation (NSF) under Grant SWIFT-2229563, the SUNY Innovative Instruction Technology Grant (IITG) 880031-03, and the U.S. Air Force Research Laboratory under Contracts FA8750-21-F-1012 and FA8750-20-C-1021.} 
\blfootnote{Distribution A. Approved for public release: Distribution Unlimited: AFRL-2024-4725 on 27 Aug 2024.} 

Through this work we aim to initiate a grassroots campaign to democratize access to wireless research testbeds with heterogeneous hardware resources and network environment, and accelerate the establishment of a mature, open experimental ecosystem for the wireless community. The API of the new Federation Plane will be released to the community after internal testing is completed. 

\end{abstract}



\maketitle

\section{Introduction}\label{sec:intro} 
Over-the-air (OTA) experimentation is becoming increasingly important for 5G-and-beyond (5G+) and IoT research, 
including closing the sim-to-real gap for data-driven networking \cite{khairy2024challenge},
improving modeling accuracy for emergent network technologies \cite{pedersen2024survey}, 
and enhancing technological readiness of complex wireless applications \cite{rafique2024slicingsurvey}. However, validating various new technologies through rigorous testbed experiments remains challenging. One of the primary obstacles is the lack of a convenient approach to share, access, and control heterogeneous experimental resources \cite{quadar2024iottestbeds}. The design and construction of testbeds require careful consideration of configuration capabilities to support broad evaluation and comprehensive documentation to generate reliable experimentation results and justify long-term investments \cite{gomez2023survey}. Consequently, innovation can be significantly hindered by the time and financial investment in acquiring hardware, documentation, and learning common ontologies, particularly for new researchers. 

\begin{table*}[t]
\centering
\footnotesize
\caption{Comparison of UnionLabs with Existing Testbed Federation Platforms}
\begin{tabular}{|l|c|c|c|c|c|}
\hline   
\multicolumn{1} {|c|} {\textbf{Feature}} & \textbf{Fed4Fire+ \cite{demeester2017fed4fire}} & \textbf{OneLab \cite{fdida2010onelab}} & \textbf{FABRIC \cite{baldin2019fabric}}  & \textbf{FIESTA-IoT \cite{sanchez2018fiestaIOT}} & \textbf{UnionLabs} \\
\hline                 
Operating System Independent & $\times$ & $\times$ & \checkmark & $\times$ & \checkmark \\
\hline
Cross-testbed Experiment Support & $\times$ & $\times$ & \checkmark & \checkmark & \checkmark \\
\hline
Require No Public IP Addresses at Endpoints & $\times$  & $\times$ & $\times$ & \checkmark & \checkmark \\
\hline
Heterogeneous Research Applications & \checkmark & \checkmark & \checkmark & $\times$ & \checkmark \\
\hline
Federation Complexity & Moderate & Not supported & High & High & Low \\
\hline
Cloud Resource Scalability & Private - fixed & Private - fixed  & N/A & Private - fixed & Public - extensible \\
\hline
\end{tabular}
\label{tab:similarplatforms}
\end{table*}

To address these challenges as part of our \textit{UnionLabs} initiative \cite{mcmanus2024unionlabs}, in this work we design and prototype a novel 
testbed federation toolchain to provide a scalable approach for shared access to a diverse set of OTA testbeds with minimal user-side complexity. 
The goal of the UnionLabs initiative is to create a grassroots campaign to democratize access to wireless research testbeds with heterogeneous experimental resources, such as algorithm, code, computing and hardware resources, as well as physical network environments. In our previous work \cite{mcmanus2024unionlabs}, we demonstrated the the first prototype of UnionLabs. This prototype was over the AWS cloud, leveraging the extensive computing, storage, and networking tools provided by the AWS ecosystem to improve user experience through cloud automation, achieve extensive scalability, and streamline platform management. With UnionLabs, experimenters are enabled to access and conduct experiments over remote testbeds with improved experimentation flexibility and reduced user-side configuration complexity. 
 
In the current implementation of UnionLabs, federating a testbed requires significant manual configuration and backend modifications to the Federation Plane on the AWS cloud. 
%
In this work, we aim to further streamline the federation process of various testbeds by automating federation tasks such as registration, namespace association, and container management, thereby effectively reducing complexity for users. 


The main contributions of this work are as follows:
\begin{itemize}   
    \item 
    We present a novel hybrid cloud testbed federation framework that leverages the comprehensive AWS toolchain for scalability and process automation. 
    %
    This framework enables the federation of testbeds through a web portal via the SMIE module, which employs simple hardware descriptions to trigger automatically the generation of testbed contexts using the DCM module. This approach facilitates streamlined deployment of remote sessions on the edge cloud deployed at individual testbeds and the execution of single-use scripts via the ASG module, which handles dependency installation and remote host configuration for federation. 
    %

    \item 
We demonstrate that the proposed testbed federation process can significantly reduce manual configuration requirements, thus decreasing user-side complexity. To this end, we prototype the framework and integrate it into the Federation Plane of UnionLabs. Utilizing the new tools provided by the Federation Plane, we federate two testbeds with UnionLabs: a LoRa IoT testbed at the University of Utah and UB NeXT, a 5G+ testbed at the University at Buffalo. Our results show that testbed federation is streamlined into a two-step process that requires no pre-configuration of edge devices. We also outline the user-side workflow for conducting experiments on the federated testbeds.     
\end{itemize}

\noindent The remainder of this paper is organized as follows:  
In Sec.~\ref{sec:related} we compare UnionLabs with existing testbed federation frameworks.  
Sec.~\ref{sec:primer} provides an overview of the 
UnionLabs resource-sharing infrastructure.  
In Sec.~\ref{sec:frameworkdesign}, we describe the major components of the new Federation Plane, and in Sec.~\ref{sec:prototype} we 
demonstrate the automated testbed federation process.  
Finally, we discuss the limitations and future work in Sec.~\ref{sec:discussion}, and draw the main conclusions in Sec.~\ref{sec:conclusion}.

\section{Related Work}\label{sec:related}
We have identified four 
existing platforms that share a similar goal with UnionLabs:
Fed4FIRE+ \cite{demeester2017fed4fire}, OneLab \cite{fdida2010onelab}, FABRIC \cite{baldin2019fabric}, and FIESTA-IoT \cite{sanchez2018fiestaIOT}. 
While other platforms, such as the NSF PAWR platforms \cite{pawr}, serve as excellent examples of remote-access testbeds with high-quality hardware and diverse OTA capabilities, they do not support the federation of new testbeds and are therefore outside the scope of this work. Before discussing the unique aspects of UnionLabs, we first give a brief overview of each of these existing platforms. 


Fed4Fire+ \cite{demeester2017fed4fire} is a European initiative aiming to provide unified access to a variety of testbeds 
including 
fixed network testbeds (e.g., Virtual Wall \cite{testbedvirtualWall}, 
PlanetLab \cite{testbedplanetlabEu}),
wireless/SDR testbeds (e.g., NiTOS \cite{testbednitos}, 
imec City Lab \cite{testbedCityLab}),
and cloud service testbeds (e.g., CloudLab \cite{testbedcloudLab}, Grid5000 \cite{testbedgrid5000}), among others. 
Fed4FIRE+ employs a Slice-based Federation Architecture (SFA), which organizes resources from multiple testbeds into user-defined "slices". Users can access these slices via a web portal, and tools such as NEPI (Network Experiment Programming Interface), SFI (Slice Federation Interface), or jFed may be required to conduct experiments in slices.
OneLab \cite{fdida2010onelab} is another international testbed federation framework initiated by the Europe.
%
%
%
The federation of testbeds is accomplished through registration with the OneLab API and the installation of three software components to connect devices to the OneLab private cloud: SFAWrap (control plane), NEPI, and TopHat (monitoring plane). At the time of writing, the OneLab platform has been federated under the Fed4FIRE+ platform, along with its constituent testbeds, and the original OneLab website does not indicate support for federation of new testbeds any longer.


\begin{figure*}[t]
    \centering
    \vspace{-4mm}
    \includegraphics[width=0.8\textwidth]{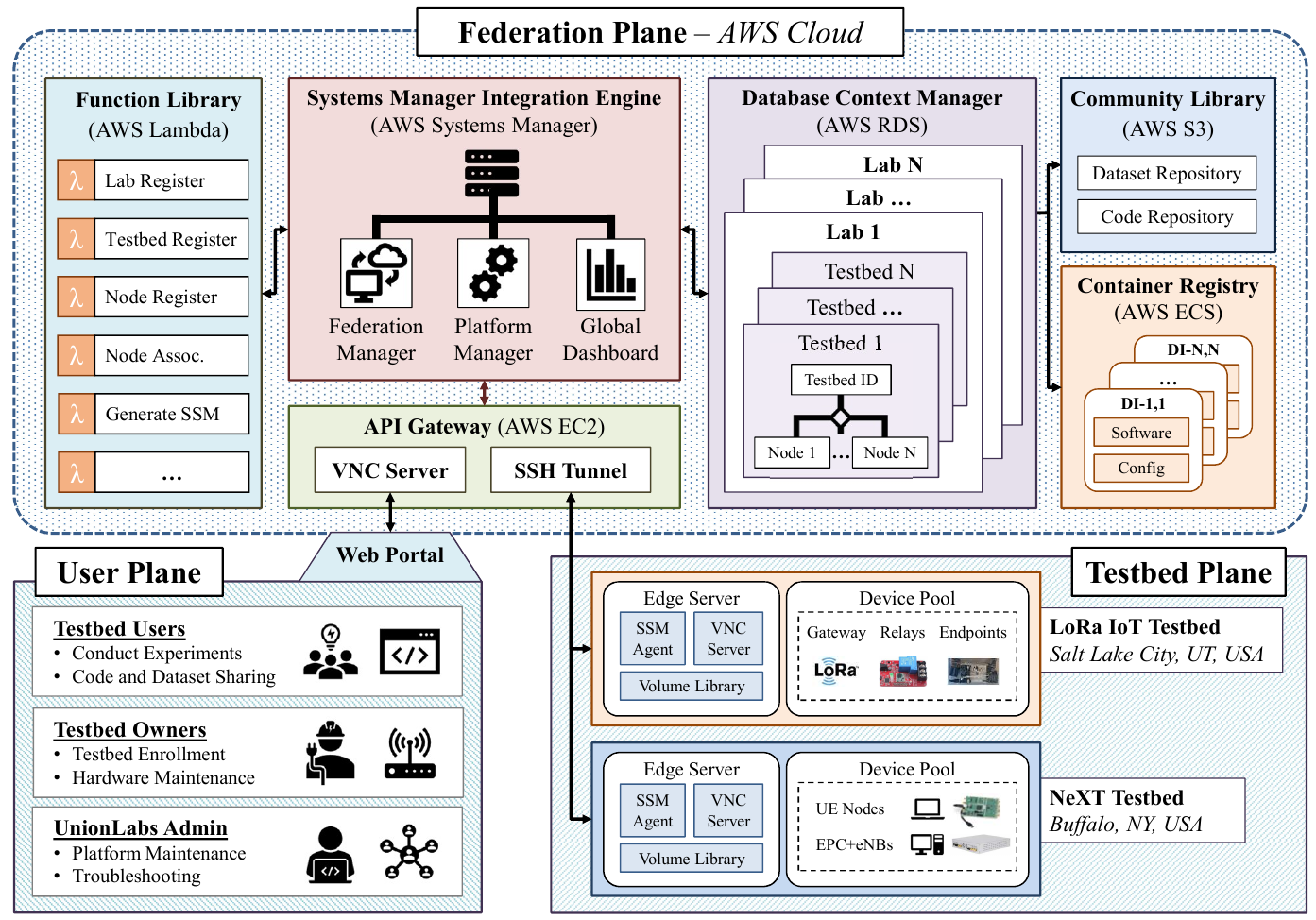}
    \caption{\small Architecture Overview of UnionLabs for Experimental Resource Sharing} \vspace{-3mm}
    \Description{UnionLabs architecture.}
    \label{fig:unionlabs_cloud}
\end{figure*}

Similar to Fed4FIRE+, FABRIC also employs an SFA architecture \cite{baldin2019fabric}. FABRIC is a U.S.-led initiative featuring a large-scale, distributed infrastructure for programmable network research. 
FABRIC consists of 40 interconnected hardware deployment sites located in Japan, the USA, the United Kingdom, Switzerland, and the Netherlands. Each site typically includes a primary control node along with various types of worker nodes, depending on network and power constraints. Access to the sites is facilitated through a fixed network of point-to-point connections. Experiments can be conducted using pre-configured virtual machines, enabling user-defined data generation through programmable layer-based protocols.
%

Finally, FIESTA-IoT \cite{sanchez2018fiestaIOT} is a federation of IoT testbeds, providing access to diverse contexts for data collection. As an "Experimentation-as-a-Service" platform, FIESTA-IoT connects users with a wide array of IoT testbeds.
The platform emphasizes semantic interoperability, employing a common ontology for experimentation, data collection, and testbed design to enhance the generality of results. Within FIESTA-IoT, testbeds are federated into a private cloud registry via a web portal and must meet five criteria for approval: usefulness, redundancy, sustainability, technical competence, and feedback. 

\textbf{Limitations of existing platforms:} 
While these platforms support open access to a diverse set of testbed resources, we have identified several key limitations. 
First, these platforms are not inherently scalable since they rely on private cloud resources, which are constrained to local hardware capabilities. 
Second, the federation process is complicated, and in some cases requires contacting platform administrators, who may even deny the federation request. 
Due to these limitations, the number of available testbeds on these platforms is insufficient to address the challenges outlined in Sec.~\ref{sec:intro}. 
For example, of the 35 testbeds federated to Fed4FIRE+, 29 were experiencing connectivity issues at the time of writing, 13 of which have been offline for at least 31 days. This leaves only six testbeds available to users. 
Further, OneLab no longer supports the federation of new testbeds. 
%
FABRIC imposes stringent hardware and network configuration requirements, potentially excluding testbed owners with limited funding or facilities from hosting deployments. 
%
Table~\ref{tab:similarplatforms} provides a more comprehensive comparison between UnionLabs and these 
platforms, in terms of federation complexity, scalability, and other aspects.  
%
To the best of our knowledge, 
\textit{UnionLabs offers the lowest complexity and the best scalability among all existing testbed federation frameworks, which we believe will benefit both testbed owners looking to federate and share experimental resources and the researchers utilizing these resources.}


\vspace{-3mm}
\section{UnionLabs: A Primer}\label{sec:primer}

As described in \cite{mcmanus2024unionlabs}, UnionLabs is a 
grassroots-driven testbed federation platform that utilizes AWS services to facilitate remote experimentation across a variety of NextG and IoT testbeds.
As illustrated in Fig.~\ref{fig:unionlabs_cloud}, the UnionLabs architecture comprises three key components: the \textit{User Plane}, the \textit{Testbed Plane}, and the \textit{Federation Plane}. The User Plane includes testbed users who conduct experiments on the platform, testbed owners who host federated hardware, and UnionLabs administrators who provide platform maintenance and general user support. The Testbed Plane encompasses all federated devices at the cloud edge, primarily RF hardware such as SDRs and IoT devices, along with associated controlling hosts like laptops or servers. The Federation Plane serves as the bridge between the User Plane and Testbed Plane, enabling real-time access to edge devices, centralized cloud monitoring and management, and storage for shared user files.

Specifically, the Federation Plane provides the following services to the User Plane and Testbed Plane: 


\begin{itemize}
\item  \textit{API Gateway}. 
%
The API Gateway connects the User Plane to testbed resources, documentation, and resource schedules through a web portal. It is deployed on a single AWS EC2 instance, which manages all user traffic and offers scalability in terms of the number of supported users and available devices.

\item \textit{Container Registry}. 
The Container Registry is a repository of Docker images associated with each edge device. These images contain all the necessary software dependencies specified by testbed owners. When a remote connection is initiated, containers are created from these images and deployed on the edge devices, ensuring consistent and efficient operation.


\item \textit{Dataset Repository}. 
This repository contains user-provided datasets that facilitate both online and offline experimentation. Users are encouraged to upload well-labeled datasets along with comprehensive documentation detailing the experimental context in which the data was generated.

\item \textit{Code Repository}. 
This code library serves as a resource for both novice experimenters looking to familiarize themselves with testbeds and experienced users aiming to design and execute reliable, repeatable experiments. Uploaded files are accompanied by descriptors that specify the relevant node, testbed, and experiment contexts, ensuring easy identification and application.
\end{itemize}

%
%
\noindent 
Based on the previous implementation of UnionLabs, to federate a new testbed, the testbed owner first collaborates with the UnionLabs administrator to develop a custom Docker image for each testbed control interface containing all necessary dependencies for testbed operation. The UnionLabs administrator then configures SSH access from each Docker container to API Gateway hosted on the AWS cloud. Once the testbed devices are accessible via the AWS cloud, a new profile is manually configured for the testbed, which contains key contexts for backend association processes, such as port assignments, Docker volume and image management, and descriptors for visibility via the web portal. This profile is then added to the database of federated resources, and linked with the corresponding namespace in the Container Registry to support deployment of the custom Docker images during the 
remote session startup process. Namespaces for this testbed are then created in the Code and Dataset Repositories and linked to this profile. Finally, the profile is used to generate a reservation module, allowing users to schedule access to these newly federated resources through the web portal.

This previous federation process required manual configuration of the end devices, local networks, and the UnionLabs backend. In Sec.~\ref{sec:frameworkdesign}, we introduce a new design for the Federation Plane that aims to automate the federation process, which has traditionally been tedious and labor-intensive.

\section{Automated Federation Framework}\label{sec:frameworkdesign}
%
As discussed in Sec.~\ref{sec:intro}, the complexity of the testbed federation process poses significant challenges to scalability and long-term growth for both testbed owners and experimenters. 
UnionLabs aims to address these challenges by providing a hybrid cloud resource-sharing infrastructure. 
%
%
%
To this end, in this section we present a toolchain to streamline the federation process
and enable testbed owners to federate their experimental resources through the web portal directly with  
\textit{zero} manual modifications of the UnionLabs backend on the AWS cloud. 
%

\begin{table}[t]
\footnotesize
 \centering
 \caption{\small Lambda Functions Designed for Federation Automation \vspace{-2mm}}
\begin{tabular}{|p{0.3\linewidth} | p{0.55\linewidth}|}
\hline   
\multicolumn{1} {|c|} {\textbf{Function Name}} & \multicolumn{1} {|c|} {\textbf{Description}} \\
\hline              
\multirow{2}{*}{Node Register} & Add private node key to global database; send activation to SSM Generator \\
\hline
\multirow{2}{*}{Testbed Register} & Add testbed group to global database by private ID; update Web Portal \\
\hline
\multirow{2}{*}{Lab Register} & Add lab group ID to global database by private ID; update Web Portal \\
\hline
\multirow{2}{*}{Node Association} & Associate node ID with testbed and lab groups; update Web Portal \\
\hline
\multirow{2}{*}{SSM Generator} & Create integration script using node key hybrid activation \\
\hline
\multirow{2}{*}{Schedule Manager}  & Check validity of new reservations and maintain master resource schedule \\
\hline
\multirow{2}{*}{Authentication} & Compare user credentials against stored profiles in global database during login \\ 
\hline
\end{tabular}
\label{tab:lambdafunctions}
\vspace{-5mm}
\end{table}

\setcounter{figure}{2}
\begin{figure*}[b]
\vspace{-2mm}
\begin{tabular}{cc}
\includegraphics[width=0.38\textwidth]{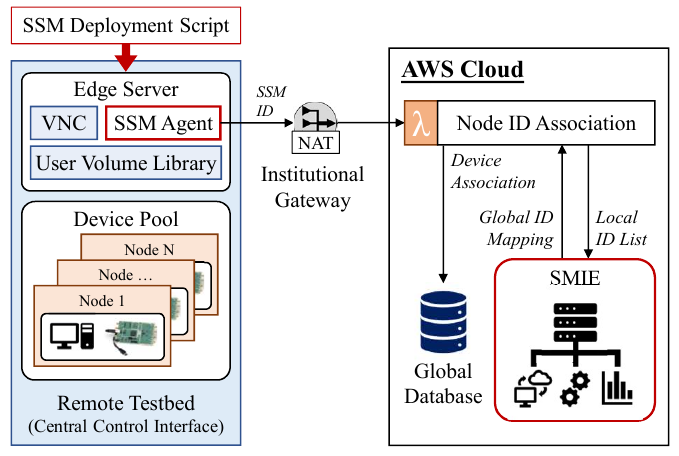} & \includegraphics[width=0.40\textwidth]{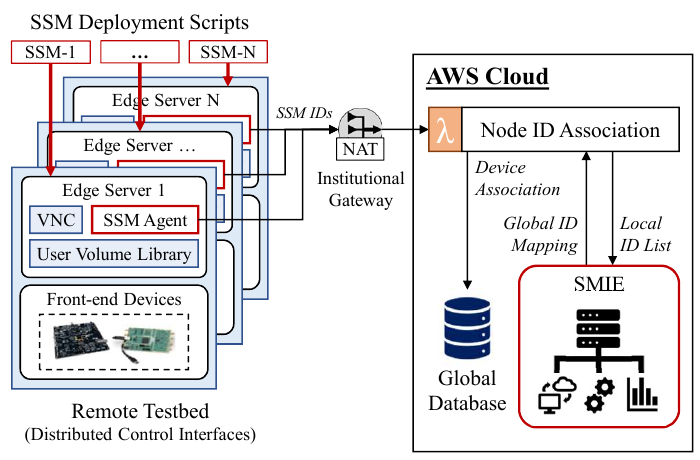} \\
\small (a) &\small \hspace{0mm} (b) \\
\end{tabular}
\vspace{-3mm}
\caption{\small 
SSM installation on testbed with (a) centralized node control interface and (b) distributed and/or direct node control interface.}
\Description{Testbed federation process.}
\label{fig:testbed_fed}
\end{figure*}  

\subsection{Toolchain Design}\label{sec:newtool}
To automate the testbed federation process, 
we first design three new tools leveraging the services provided by the AWS platform management framework such as the Systems Manager, Relational Database Service (RDS), and EC2. These new tools are \textit{Function Library}, \textit{System Manager Integration Engine}, and \textit{Database Context Manager}. 
Then, we integrate these tools to the Federation Plane deployed on the AWS cloud via a library of AWS Lambda functions. Next, we describe these three tools sequentially. 


\textit{Function Library}. 
We design the Function Library based on AWS Lambda, an event-driven computing service that enables the execution of user-defined code in response to specific events or state changes, such as user requests, database updates, or the activation and completion of other functions. The library comprises a collection of AWS Lambda functions developed to facilitate automated interactions among Federation Plane services, including managing backend control sequences, script generation, and data exchanges triggered by user interactions. These functions have been deployed across the Federation Plane enabling UnionLabs to abstract the complexity of federation operations from both testbed users and owners, thereby enhancing the user experience. Services automated by this library include registration of labs, testbeds, and associated devices, the definition of lab and testbed namespaces in accordance with the UnionLabs ontology, and the automatic interpolation of data necessary for device federation. A detailed summary of these Lambda functions is provided in Table~\ref{tab:lambdafunctions}.

\textit{Systems Manager Integration Engine (SMIE)}. 
The SMIE is designed to centralize monitoring and backend control for all federated devices leveraging AWS System Manager tools. 
Among the extensive features offered by AWS Systems Manager, we have identified three categories that can simplify user experience and enhance scalability for testbed federation:
i) Federation Manager: Comprising the Hybrid Activation and Run Command features, this component initiates backend federation processes, including federation script creation, AWS Service Manager Agent (SSM Agent) deployment, and database maintenance, based on data provided by the API Gateway;
ii) Platform Manager: Utilizing the Fleet Manager and Run Command features, the Platform Manager facilitates testbed access for both user experimentation and platform maintenance by UnionLabs administrators; and iii) Global Dashboard: This feature provides a centralized, global overview of all federated devices by leveraging the Fleet Manager. It includes details on device hardware and software status, private and public identifiers, and network-level information necessary for maintaining remote access capabilities.
The federated devices interface with the Systems Manager through the deployment of SSMs, which enable direct cloud services engagement with testbed resources. These agents are configured by the SMIE to ensure seamless interoperability with other Federation Plane services, supporting the integration of new features and quality-of-service (QoS) updates throughout the device lifecycle.



\textit{Database Context Manager (DCM)}.
The DCM is the primary data management service designed for the UnionLabs Federation Plane. It automatically generates mutual associations among all resources, testbeds, labs, and experiments. Interfaced with AWS Relational Database Service (RDS), the DCM provides mapping services for federated experimental resources at various levels, including node, testbed, and experiment contexts. These mappings are utilized by other Federation Plane modules, such as the Service Manager and Lambda functions, to orchestrate reliable automation of user interactions and testbed organization. Additionally, the DCM identifies which devices are associated with each testbed on the web portal and presents documented experiments contained in the Code or Dataset Repositories related to the reserved experimental resources.

\setcounter{figure}{1}
\begin{figure}[t]
    \centering
    \includegraphics[width=0.45\textwidth]{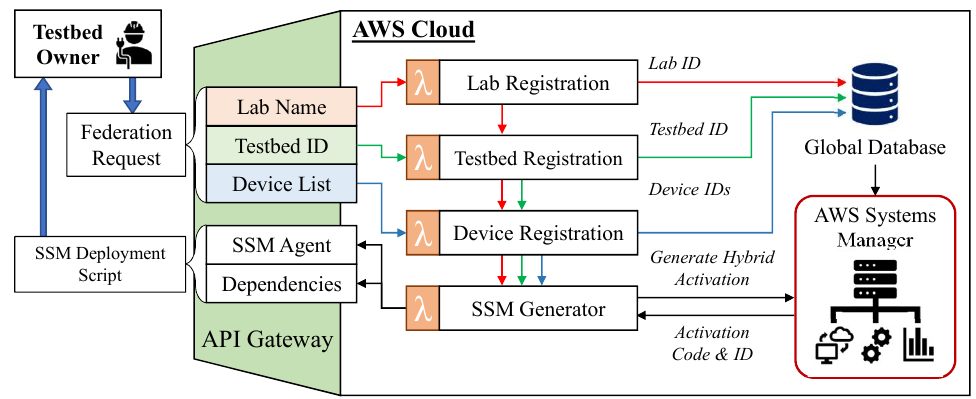}
    \caption{\small Automated SSM script generation for testbed integration.}
    \Description{Testbed federation script generation.}\vspace{-5mm}
    \label{fig:testbed_fed_1}
\end{figure}

\subsection{Federation Process} \label{sec:fedprocess}
%
%
With the new tools designed in Sec.~\ref{sec:newtool},  federating a testbed can be streamlined into a simple two-step process.  
As discussed in Sec.~\ref{sec:related}, this process can be completed by testbed owners through the UnionLabs web portal, without administrative assistance. Therefore, this approach can significantly reduces the complexity for testbed owners and encourages a wide range of prospective researchers to integrate their hardware with the platform.
The first step involves registering all testbed control interfaces, such as laptops or servers, through the User Dashboard on the UnionLabs web portal. As depicted in Fig.~\ref{fig:testbed_fed_1}, the registration process begins when the testbed owner submits an integration request, which includes the laboratory name, a public testbed identifier, an overview of the testbed's general purpose and key capabilities, and a list of public identifiers for each testbed control interface, along with detailed descriptions of connected devices. Once the integration request is submitted, a unique internal ID is automatically generated for each new device, which will be used to map new hardware contexts to all associated backend processes. This ensures that the testbed becomes visible under the associated lab group on the web portal and that each federated device is displayed within the testbed entry. Then, the Systems Manager Integration Engine (SMIE) module (see Sec.~\ref{sec:newtool}) uses these new contexts to generate a Hybrid Activation for each registered control interface, as illustrated in Fig.~\ref{fig:testbed_fed_1}. The output of this process is a key-value pair that links the unique internal device ID with a single-use SSM deployment script specific to that device. This script, created by the 
Automated Script Generator (ASG),
consolidates all necessary steps for deploying the UnionLabs software environment, including all required dependencies and configuration steps, into a single terminal command. An example of the 
script will be given in Sec.~\ref{sec:prototype}.

The second step of the federation process is to simply copy the generated SSM deployment script and run it on the control interface intended for federation. 
This procedure is consistent across different testbeds but may vary in intensity depending on the configuration of testbed control interfaces and peripheral hardware. For testbeds with a centralized control interface, such as distributed wireless sensor networks or cloud testbeds, testbed owners will receive a single SSM deployment script to run on the central interface. For instance, as illustrated in Fig.~\ref{fig:testbed_fed}(a), if there is central node controlling all sensor devices in the testbed, only that control node will need to be registered. In contrast, for testbeds with distributed control, such as large-scale SDR testbeds, each node control interface must be registered separately and will receive a unique SSM deployment script, as shown in Fig.~\ref{fig:testbed_fed}(b). Regardless of the configuration, executing this script on each interface will deploy the SSM and allow the Database Context Manager (DCM) to generate an internal context in the global database and link it with backend services, thereby extending the control and monitoring capabilities of the SMIE to that device .




\setcounter{figure}{3}
\begin{figure}[t]
    \centering
    \includegraphics[width=0.33\textwidth]{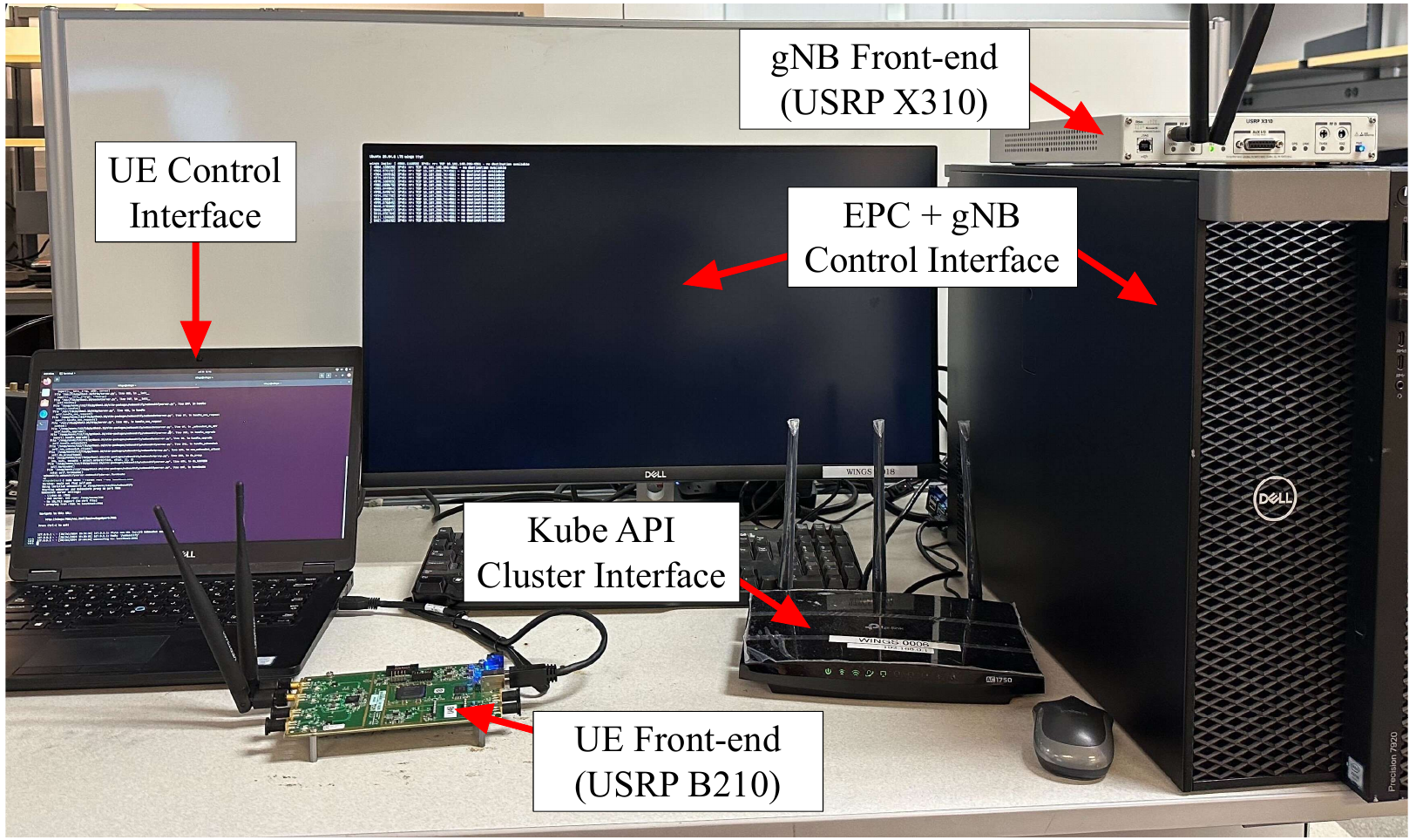}
    \vspace{-4mm}
    \caption{\small Part of the resources available to the UB NeXT testbed. \vspace{-4mm}} 
    \label{fig:next_snapshot}
\end{figure}

\vspace{-3mm}
\subsection{Service Integration}
To enable seamless integration with automation services via shared hardware contexts, in addition to the new tools introduced above, we have also redesigned most of the core modules of the Federation Plane outlined in Sec.~\ref{sec:primer}, leveraging the AWS toolchain. 
For example, the Container Registry, previously deployed and maintained locally on edge cloud at the testbed side using Ansible for Docker container creation and deployment, is now powered by the AWS Elastic Container Service (ECS). 
ECS's ability to associate with device contexts generated by the DCM enables direct interoperability with the SMIE, streamlining and accelerating remote session initialization.
Additionally, the Code Repository and Dataset Repository, which were initially implemented on local server storage requiring manual updates for new devices, testbeds, and experimental contexts, are now hosted in the public cloud using AWS S3. This transition allows uploaded files to automatically link with existing DCM-generated contexts in the Global Database, significantly expediting the publication process.
Furthermore, migrating to the AWS toolchain provides enhanced global security through AWS Cognito, which manages user authorization, access control, and data security in the public cloud.

\setcounter{figure}{3}

\section{Demonstration and Evaluation}\label{sec:prototype}
We have prototyped the federation framework described in Sec.~\ref{sec:frameworkdesign} using the AWS cloud and integrated it to the Federation Plane of UnionLabs. 
In this section, we first demonstrate the federation framework considering specific wireless testbed, 
then we provide an overview of the framework's performance considering metrics that impact the user-side testbed federation experience and the initiation of remote access sessions.

\subsection{Federation Demonstration}
We demonstrate the federation process taking the UB NeXT testbed as an example.
NeXT is a software-defined wireless \textbf{Ne}twork \textbf{X}-Control \textbf{T}estbed developed in our previous research for 5G+ research, with variable “X” referring to optimization, simulation and experimentation \cite{hu2023nextjournal}. 
Figure~\ref{fig:next_snapshot} shows part of the resources available to the NeXT testbed, 
including UE (USRP B210), core network (EPC+gNB: USRP X310), and Kube API Cluster Interface. 
%
%
%
\setcounter{figure}{4}
\begin{figure}[t]
    \centering
    \includegraphics[width=0.43\textwidth]{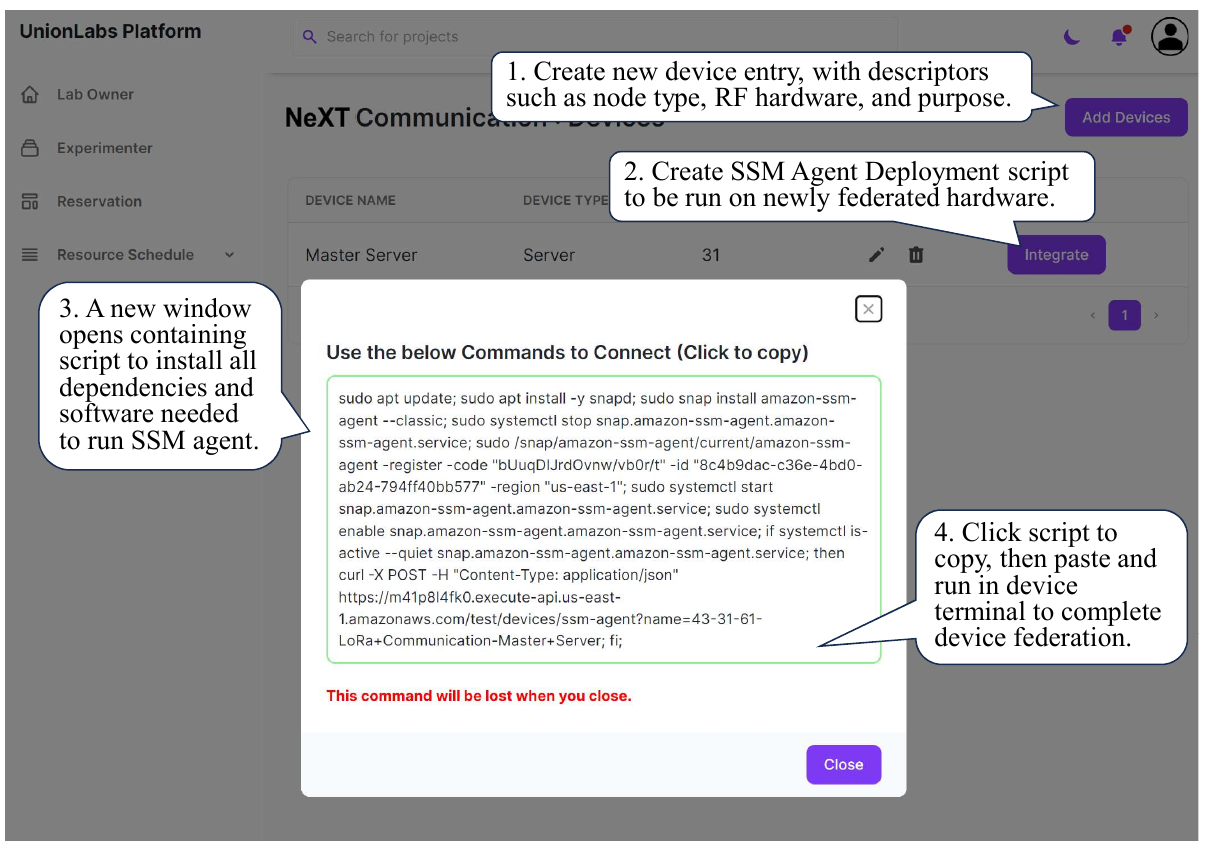}
    \vspace{-4mm}
    \caption{\small Example of auto-generation of SSM installation script. 
    }
    \vspace{-8mm}
    \Description{Testbed federation script example.}
    \label{fig:testbed_fed_screenshot}
\end{figure}

\begin{figure*}[t]
    \centering
    \includegraphics[width=0.7\textwidth]{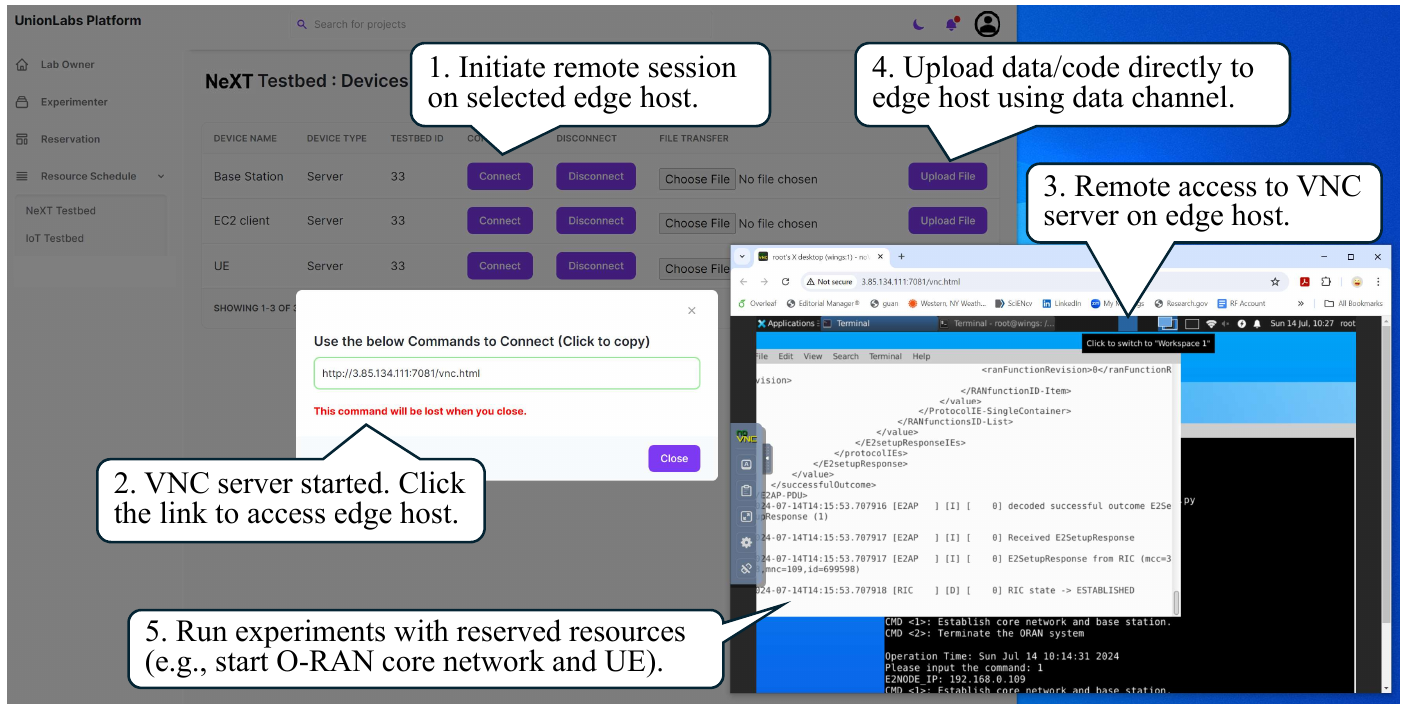}
    \vspace{-4mm}
    \caption{\small Screenshot of remote experiment on the federated NeXT testbed.} \vspace{-3mm}
    \Description{Remote experiment screenshot.}
    \label{fig:oran_epc_enb}
\end{figure*}

Before federating the testbed, we first create a blank profile called "NeXT Communication Testbed", which can be done by simply clicking a button within the UnionLabs web portal. 
This creates a table in the API Gateway which is used to link user commands issued through the web portal, such as resource reservation, device integration, and remote access initialization commands, to the appropriate backend services via Lambda functions (see Sec.~\ref{sec:frameworkdesign}).

Then, we can add new device entries on the automatically created testbed page through "Add Devices", as shown in Fig.~\ref{fig:testbed_fed_screenshot} (Step~1).
In this example, the edge computing resources of the testbed is provided by its Master Server, a Dell Precision 7920 workstation with Intel Xeon Gold 5215 2.5 MHz CPU, 64 GB ($8\times8$ GB) DDR4 2933 MHz memory, running Ubuntu 20.04.6 LTS Server operating system. It serves as the controlling host of the gNB front-end (i.e., the USRP X310) as well as the bridge between the gNB and the UnionLabs Federation Plane deployed on the AWS pubic cloud. The Master Server can be federated using the "Integrate" functionality (Step~2 in Fig.~\ref{fig:testbed_fed_screenshot}). As described in Sec.~\ref{sec:fedprocess}, 
the SSM deployment script will be generated by the Automated Script Generator (ASG).
When the ``Integrate" command is triggered on the web portal, the API Gateway sends the provided hardware descriptors to the SMIE to generate a hybrid activation, which triggers a Lambda function to parse the resulting unique activation code and ID to a preconfigured SSM deployment script. 
This script first updates the package repository on the end device, then installs the \texttt{snapd} package manager. From \texttt{snapd}, the SSM software is installed and configured using the unique hybrid activation code and ID. The SSM is then activated, which updates the internal Node ID flag in the DCM via Lambda function, indicating successful device federation. 
Once completed, the DCM then creates a namespace in the Global Database which associates the new Node ID with various contexts stored throughout the framework such as testbed- or device-specific software containers in the Container Registry, user files stored in the Code and Dataset Repositories, and user commands issued via the API Gateway.

Once the testbed is federated, a resource reservation page will be created automatically.
Through this page, 
the experimenter is allowed to reserve the hardware and computing resources available to the testbed, specifying selected resources, date, and duration of experiment, or access available resources for immediate use. 
As mentioned in Sec.~\ref{sec:frameworkdesign}, resources can be federated through centralized or distributed control interfaces. In the case of centralized control interfaces, 
%
resources can be reserved and accessed through the master server, which may provide endogenous control paths between the master server and any other edge nodes or control interfaces, or else direct access to experimental devices, such as SDRs, connected directly to the master server.
In the case of distributed control interfaces, individual nodes must be reserved and accessed from the testbed page on the web portal, which generally allows the same level of control over any experimental devices connected to that node as would be available if connected directly to local machine, i.e. GNURadio or srsRAN, for example, according to the Docker container deployed on that node. 
In either case, additional experimental hardware can be incorporated by testbed owners by connecting them to federated control interfaces, or federating another control interface for them following the same process.
%

The federation process, including logging on to the web portal, generating descriptors, and receiving the SSM deployment script (Fig.~\ref{fig:testbed_fed_screenshot}), takes a total of only two minutes. For testbeds with multiple devices, those common descriptors such as RF hardware specifications, testbed and laboratory identifiers, and hardware capabilities can be reused among devices with similar capabilities to further accelerate the testbed federation process. 


\vspace{-3mm}
\subsection{Remote Experiment Demonstration}
\begin{figure}[t]
    \centering
    \includegraphics[width=0.4\textwidth]{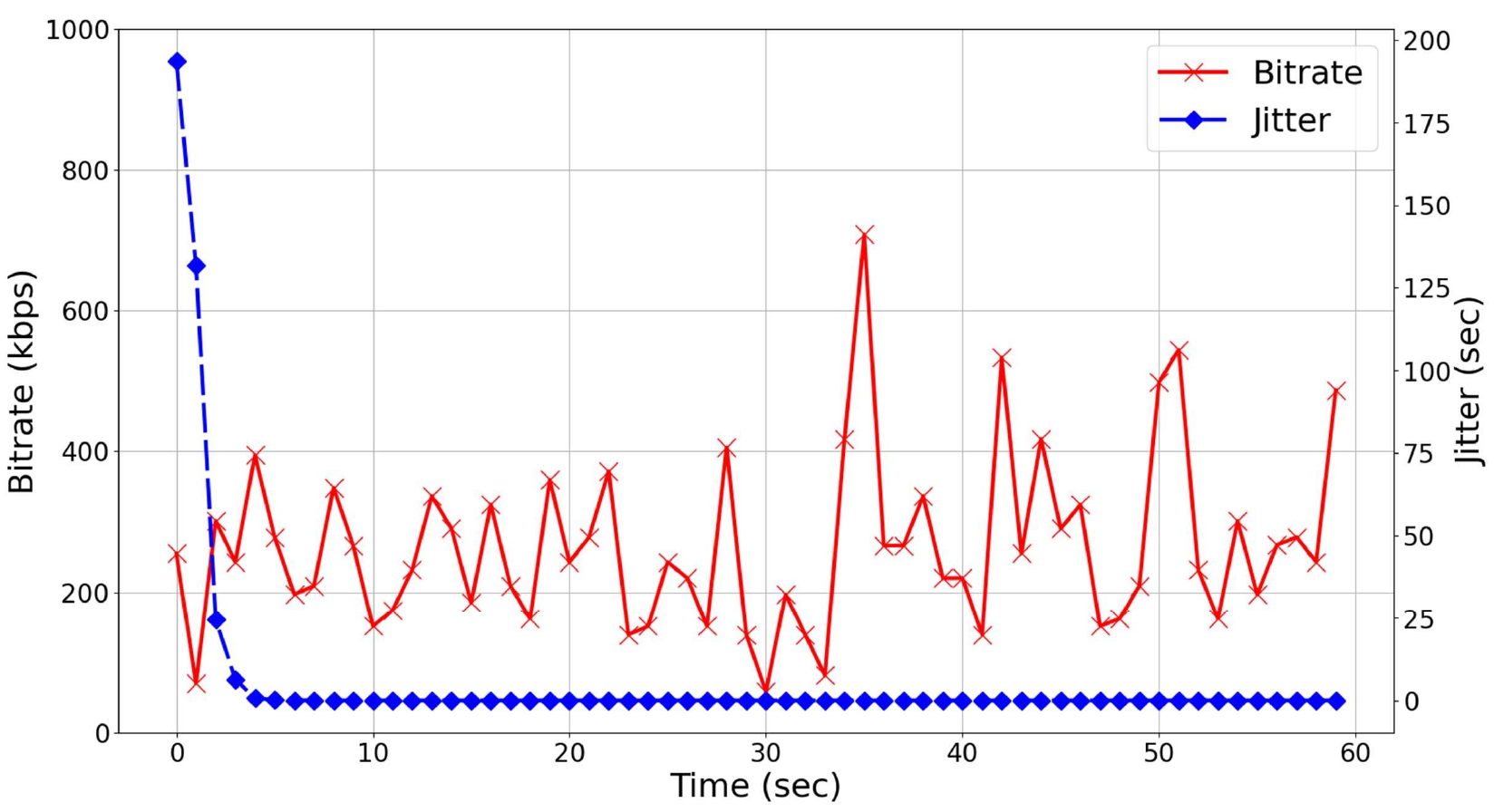}
    \vspace{-2mm}
    \caption{\small Results of rate and jitter measurement for uplink communication between UE and gNB.}\vspace{-2mm}
    \Description{Experimental results.}
    \label{fig:oran_speedtest}
\end{figure}

We showcase the remote experimentation considering a toy example. In this experiment, we will establish an O-RAN uplink between the UE and gNB. To this end, as shown in Fig.~\ref{fig:oran_epc_enb} (Step 1), we can 
remotely access the hardware resources available to the federated NeXT testbed 
using the ``Connect" functionality. 
This will deploy a Docker container created by the testbed owner which has been associated with this node in the Global Database, which installs the OAIC protocol stack and Near-Real Time RAN Intelligent Controller (RIC) \cite{oaicGithub} on this node, and starts a VNC server for remote access and control of the Master Server. 
Once this process is finished, the user is provided a link which will open a new browser window on the experimenter's local device connected to that server (Step 2 in Fig.~\ref{fig:oran_epc_enb}). 
At this point, we are allowed to access and control various computing and SDR resources interfaced to the server. In this example,  
we are connected to a VNC server on an edge node of the NeXT testbed. 
While connected, we are able to upload code, datasets, or other experiment files for use on this node via the web portal. 
Finally, we start the O-RAN core network on this node and a UE on another node to conduct OTA experiments using the reserved hardware. 
%
%
Following this procedure, we have connected to the NeXT testbed node and started the core network (EPC + gNB) and the UE processes. 
%
Once the network is set up and the UE is attached to the core network, we run \texttt{iperf} to measure data rate and jitter for the uplink data path. 
The results are shown in Fig.~\ref{fig:oran_speedtest}, in which we plot the jitter of a low-data rate (avg. 260 kbps) link between the two nodes.

\subsection{Cloud-to-Edge Latency Performance}
UnionLabs has been designed following a cloud-edge resource sharing infrastructure. 
To assess the effectiveness of this architecture, we further evaluate the node access latency between 
AWS cloud and edge servers, defined as the duration from the initiation of remote access through the web portal until the testbed edge server interface is visible. 
To this end, in addition to the UB NeXT testbed, we have also federated a testbed deployed at the University of Utah in Salt Lake City, UT, which specializes in IoT research. Due to space limitations, details of the IoT testbed have been omitted. 
%
The average, minimum, and maximum latency for remote access instance deployment
are shown in Table~\ref{tab:accesstime}. 
It can be seen that the new, 
automated hybrid cloud architecture achieves a substantial improvement in node access speed, averaging 189\% faster than the previous platform design. This enhancement provides a significant advantage for rapid experimentation and a seamless user experience. It is worth noting that the latency associated with manual federation does not account for the time required to modify the UnionLabs backend when federating the testbed, which can range from several hours to several days.

\begin{table}[t]
 \centering
 \footnotesize
     \caption{Comparison of Node Access Latency}
\begin{tabular}{|l|c|c|c|}
\hline   
\multirow{2}{*}{\textbf{Cloud Architecture}} & \multicolumn{3}{c|}{\textbf{Latency (sec)}} \\
\cline{2-4}                 
& \textbf{Average} & \textbf{Maximum} & \textbf{Minimum} \\
\hline
Manual Federation & 60.58 & 75.97 & 51.60 \\
\hline
Automated Federation & \textbf{11.47} & 12.59 & 9.67 \\
\hline
\end{tabular}
\label{tab:accesstime} \vspace{-3mm}
\end{table}

\begin{figure}[t]
    \centering
    \includegraphics[width=0.35\textwidth]{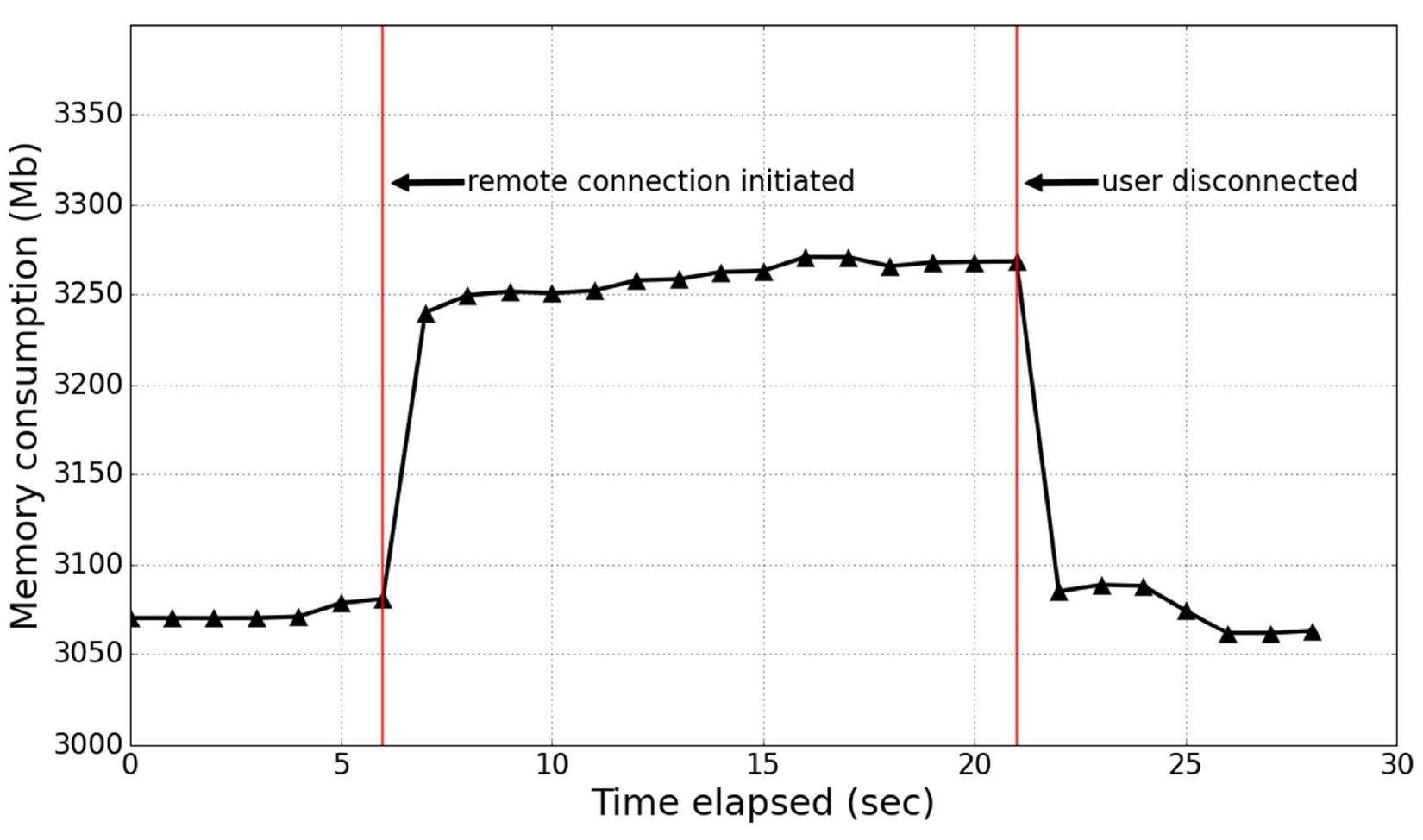}
    \caption{\small Memory consumption of SSM during remote access. 
    } \vspace{-3mm}
    \Description{SSM memory footprint.}
    \label{fig:ssm_memory}
\end{figure}

In order to help testbed owners estimate the resource needs of federated hardware and assess expected performance, we further evaluate the hardware requirements for federated edge hosts by measuring the memory consumption of the SSM processes during remote access via the web portal.  As depicted in Fig.~\ref{fig:ssm_memory}, the average memory usage of the SSM and its associated processes during remote access is approximately 200 MB. In contrast, background processes on the selected edge server within the O-RAN testbed consume roughly 3.07 GB of memory. Additionally, our assessment of CPU utilization reveals no significant increase due to remote edge server access. These findings suggest that the resources required for SSM operation on edge devices remain minimal even during remote operations, thereby alleviating performance constraints on federated hardware. This outcome supports the goal of lowering barriers for testbed owners seeking to federate their testbeds, particularly in cases involving resource-constrained hardware such as wireless sensor devices or flying vehicles.

\vspace{-3mm}
\section{Limitations and Future Work}\label{sec:discussion} 
We believe that our work on UnionLabs can help democratize access to wireless research testbeds with heterogeneous hardware resources and network environment, and accelerate the establishment of a mature, open experimental ecosystem for the wireless community. We acknowledge several limitations, which will be addressed in future work.

    $\bullet$ \textbf{Support of mobile nodes.} 
    Current observations of remote testbeds are confined to a VNC server display on each node, which restricts UnionLabs' capability to support multi-domain testbeds, such as those for aerial networking, where visualization of node mobility is crucial. In future developments, we plan to create tools that will enable researchers to stream live video of experiments and interact with mobile nodes in real-time through the Web Portal.
    
   $\bullet$ \textbf{Automated container coordination.} Currently, the platform only supports the use of Docker for software environment containerization. This may be a limiting factor for testbed owners with no Docker experience. To address this, we are working to deploy Kubernetes Clusters as part of the federation process, which will include a Control Plane and a Worker Node, or "pod". The Control Plane will be interfaced with the AWS Elastic Kubernetes Service for centralized monitoring and streamlined integration with the rest of the UnionLabs ecosystem. The pod supports several container runtime formats, which removes the strict reliance on Docker and enables compatibility with many container runtime environments. 
    
    $\bullet$ \textbf{Administrative portal development.} 
    We also plan to develop an Administrative Portal to serve as a comprehensive dashboard for managing user experience and service reliability. This portal will provide real-time monitoring of system performance and resource utilization, including device health, spectrum usage, and overall resource statistics. Additionally, it will feature automated alerts for anomalous or erroneous behavior of both UnionLabs and its users, facilitating proactive platform maintenance and enhancing transparency.    


\section{Conclusion}\label{sec:conclusion}

In this work, we have presented a new federation framework for UnionLabs, a cloud-based infrastructure aiming to democratize access to wireless research testbeds and foster an open experimental ecosystem for the wireless community. This framework simplifies federation for testbed developers by automating backend operations, enabling scalable federation and remote access to wireless testbeds. Key components such as SMIE, ASG, and DCM were developed to support this framework. We deployed a prototype of the Federation Plane on AWS and demonstrated its effectiveness by federating two testbeds: UB NeXT at the University at Buffalo and UT IoT at the University of Utah. In future work, we plan to develop tools to support federation of testbeds for 5G+ and IoT research with mobile nodes. 
\bibliographystyle{ieeetr}
\bibliography{unionlabs}

\begin{thebibliography}{10}

\bibitem{khairy2024challenge}
S.~Khairy, G.~Mittag, V.~Gopal, F.~Yan, Z.~Niu, E.~Ameri, S.~Inglis, M.~Golestaneh, and R.~Cutler, ``{Bandwidth Estimation in Real Time Communications Challenge},'' in {\em Proc. of ACM Multimedia Systems Conference}, (Bari, Italy), April 2024.

\bibitem{pedersen2024survey}
K.~Pedersen, R.~Maldonado, G.~Pocovi, E.~Jian, M.~Lauridsen, I.~Vov{\'a}cs, M.~Brix, and J.~Wigard, ``{A Tutorial on Radio System-Level Simulations With Emphasis on 3GPP 5G-Advanced and Beyond},'' {\em IEEE Communications Surveys and Tutorials (Early Access)}, pp.~1--35, March 2024.

\bibitem{rafique2024slicingsurvey}
W.~Rafique, J.~Barai, A.~Fapojuwo, and D.~Krishnamurthy, ``{A Survey on Beyond 5G Network Slicing for Smart Cities Applications},'' {\em IEEE Communications Surveys and Tutorials (Early Access)}, pp.~1--35, June 2024.

\bibitem{quadar2024iottestbeds}
N.~Quadar, M.~Rahouti, M.~Ayyash, S.~Jagatheesaperumal, and A.~Chehri, ``{IoT-AI}/machine learning experimental testbeds: The missing piece,'' {\em IEEE Internet of Things Magazine}, vol.~7, pp.~136--143, January 2024.

\bibitem{gomez2023survey}
J.~Gomez, F.~Hfoury, J.~Crichigno, and G.~Srivastava, ``{A survey on network simulators, emulators, and testbeds used for research and education},'' {\em Computer Networks}, vol.~237, p.~110054, Dec. 2023.

\bibitem{demeester2017fed4fire}
P.~Demeester, P.~V. Daele, T.~Wauters, and H.~Hrasnica, ``{{Fed4FIRE} - The Largest Federation of Testbeds in Europe},'' in {\em Building the Future Internet through {FIRE}} (M.~Serrano, N.~Isaris, and H.~Schaffers, eds.), pp.~87--109, New York, NY, USA: River Publishers, 1~ed., 2017.

\bibitem{fdida2010onelab}
S.~Fdida, T.~Friedman, and T.~Parmentelat, ``{{OneLab} - An Open Federated Facility for Experimentally Driven Future Internet Research},'' in {\em New Network Architectures: The Path to the Future Internet} (T.~Tronco, ed.), pp.~141--152, Heidelberg, Germany: Springer Berlin Heidelberg, 1~ed., 2010.

\bibitem{baldin2019fabric}
I.~Baldin, A.~Nikolich, J.~Griffioen, I.~Monga, K.~Wang, T.~Lehman, and P.~Ruth, ``{FABRIC: A national-scale programmable experimental network infrastructure},'' {\em IEEE Internet Computing}, vol.~23, pp.~38--47, Nov. 2019.

\bibitem{sanchez2018fiestaIOT}
L.~S\'{a}nchez, J.~Lanza, J.~R. Santana, and R.~Agarwal, ``{Federation of Internet of Things Testbeds for the Realization of a Semantically-Enabled Multi-Domain Data Marketplace},'' {\em Sensors}, vol.~18, p.~3375, Oct. 2018.

\bibitem{mcmanus2024unionlabs}
M.~McManus, T.~Rinchen, S.~M. Suhail, S.~Santhinivas, A.~Dey, S.~Pagidimarri, Y.~Cui, J.~Hu, J.~Zhang, X.~Wang, M.~Ji, N.~Mastronarde, and Z.~Guan, ``{Demo Abstract: {UnionLabs}: {AWS}-based Remote Access and Sharing of {NextG} and {IoT} Testbeds},'' in {\em IEEE International Conference on Computer Communications (INFOCOM)}, (Vancouver, Canada), May 2024.

\bibitem{pawr}
 https://advancedwireless.org/.

\bibitem{testbedvirtualWall}
{imec iLab.t}, ``{Virtual Wall},'' https://doc.ilabt.imec.be/ilabt/virtualwall/.

\bibitem{testbedplanetlabEu}
{PlanetLab}, ``{PlanetLab Europe},'' https://www.planet-lab.eu/.

\bibitem{testbednitos}
NITLab, ``{Network Implementation Testbed Using Open Source Platforms (NITOS)},'' https://nitlab.inf.uth.gr/NITlab/nitos.

\bibitem{testbedCityLab}
{imec iLab.t}, ``{CityLab},'' https://doc.lab.cityofthings.eu/.

\bibitem{testbedcloudLab}
{The University of Utah}, ``{CloudLab},'' https://www.cloudlab.us/.

\bibitem{testbedgrid5000}
{SILECS}, ``{Grid5000},'' https://www.grid5000.fr/.

\bibitem{hu2023nextjournal}
J.~Hu, Z.~Zhao, M.~McManus, S.~K. Moorthy, Y.~Cui, N.~Mastronarde, E.~S. Bentley, M.~Medley, and Z.~Guan, ``{NeXT: Architecture, prototyping and measurement of a software-defined testing framework for integrated RF network simulation, experimentation and optimization},'' {\em Computer Communications}, vol.~210, pp.~342--355, October 2023.

\bibitem{oaicGithub}
{OAIC}, ``{Open AI Cellular},'' https://github.com/openaicellular/oaic.

\end{thebibliography}


\end{document}